\documentclass[letterpaper, 10 pt, conference]{ieeeconf}  % Comment this line out if you need a4paper

\IEEEoverridecommandlockouts                   
\usepackage{graphics} % for pdf, bitmapped graphics files
\usepackage{epsfig} % for postscript graphics files
\usepackage{mathptmx} % assumes new font selection scheme installed
\usepackage{times} % assumes new font selection scheme installed
\usepackage{amsmath} % assumes amsmath package installed
\usepackage{amssymb}  % assumes amsmath package installed
\usepackage[T1]{fontenc}
\usepackage[latin9]{inputenc}
\usepackage{amsmath}
\usepackage{graphicx}
\usepackage{authblk}
\usepackage{color}
\newcommand{\revise}[1]{\textcolor{black}{#1}}

\title{\LARGE \bf How Decentral Smart Grid Control limits non-Gaussian power grid frequency fluctuations}

\author{Benjamin Sch\"afer$^{1}$, Dirk Witthaut$^{2}$ and Marc Timme$^{1}$% <-this % stops a space
\thanks{$^{1}$Benjamin Sch\"afer and Marc Timme  are with the Chair for Network Dynamics, Center for Advancing Electronics Dresden (cfaed),the Institute for Theoretical Physics, Technical University of Dresden, Dresden and Network Dynamics, MPIDS, G\"ottingen, Germany      
        {\tt\small benjamin.schaefer@tu-dresden.de | marc.timme@tu-dresden.de}}%
\thanks{$^{2}$Dirk Witthaut is with the Forschungszentrum J\"ulich, Institute of Energy and Climate Research-Systems Analysis and Technology Evaluation, J\"ulich and the Institute for Theoretical Physics, University of Cologne, K\"oln, Germany
        {\tt\small d.witthaut@fz-juelich.de}}%
}

\begin{document}

\maketitle
\thispagestyle{empty}
\pagestyle{empty}

%%%%%%%%%%%%%%%%%%%%%%%%%%%%%%%%%%%%%%%%%%%%%%%%%%%%%%%%%%%%%%%%%%%%%%%%%%%%%%%%
\begin{abstract}
\revise{Frequency fluctuations in power grids, caused by unpredictable renewable
energy sources, consumer behavior and trading, need to be balanced
to ensure stable grid operation. Standard smart grid solutions to
mitigate large frequency excursions are based on centrally collecting
data and give rise to security and privacy concerns. Furthermore, control of fluctuations is often tested by employing Gaussian perturbations.
Here, we demonstrate
that power grid frequency fluctuations are in general non-Gaussian,
implying that large excursions are more likely than expected based
on Gaussian modeling. We consider real power grid frequency measurements from Continental Europe and compare them to stochastic models and predictions based on Fokker-Planck equations. Furthermore, we review a decentral smart grid control scheme
to limit these fluctuations. In particular, we derive a scaling law of how  decentralized control actions reduce the magnitude of frequency fluctuations and
demonstrate the power of these theoretical predictions using a test grid. Overall, we find that decentral
smart grid control may reduce grid frequency excursions due to both
Gaussian and non-Gaussian power fluctuations and thus offers an alternative
pathway for mitigating fluctuation-induced risks.
}
\end{abstract}

%%%%%%%%%%%%%%%%%%%%%%%%%%%%%%%%%%%%%%%%%%%%%%%%%%%%%%%%%%%%%%%%%%%%%%%%%%%%%%%%
\section{INTRODUCTION}

%ToDo: Revise Discussion to be clearer structured? (along the main findings of the paper
%Restructure Introduction: Motivation, Literature survey, Our approach, Structure of the paper

Electric power grids operate close to set reference frequencies (e.g.,
$f_{R}=50~\text{Hz}$) to robustly ensure power distribution among
generators and consumers \cite{Machowski2011}. During operation,
the grid is experiencing frequency fluctuations arising from power
fluctuations on the supply side, e.g. due to fluctuating feed-in from renewable energy
sources \cite{Sorensen2007,Milan2013}, as well as power fluctuations in consumption
\cite{Wood2013} and signals resulting from energy trading \cite{NationalAcademiesofSciences2016}.
Already today, frequency fluctuations can be substantial \cite{Carrasco2006,UK-Frequency2016}
which can be thretening for sensitive electronic equipment \cite{Kundur1994,Omran2011,Li2013};
in particular since large shares of renewable generation, like solar
generation, are placed spatially distributed on low voltage levels, making it difficult to control by central control schemes \cite{Boemer2011}.
In future power grids, frequency stability is expected to become a
major issue, primarily due to the replacement of synchronous machines
by power electronics, which have no natural inertia \cite{Kroposki2008}.
Furthermore, a recent analysis showed that frequency fluctuations
are anomalous, e.g. follow non-Gaussian statistics, such that large
deviations can be more likely than expected \cite{Schaefer2017a}.
Reliably operating 100\% renewable grids of the future is thus confronted
with major challenges. One open question is: How to best limit the
impact of power fluctuations on the grid frequency?

\revise{There exist several different approaches on how grids with a high share of renewable generation can be controlled \cite{Blaabjerg}. Novel concepts for frequency control include for example the  usage of  storage \cite{DeLille, Mercier}, controlling wind turbines directly \cite{Margaris} or applying  demand side control \cite{Molina-Garcia,Short}.
On the demand side, a variety of smart grid concepts \cite{Amin2005,Fang2012}, have been
proposed to balance supply-demand differences and to limit fluctuations.}
However, standard smart grid concepts are often centralized and based on collective
supply and demand information of the generator and consumer sides
in real time. 
Thereby, the instantaneous imbalance is computed centrally
and the grid should be balanced by sending incentives, e.g. price
signals, to increase or decrease supply and demand \cite{Albadi2008,Palensky2011, Martyr2018}.
Such central organization comes with a number of issues that may be
undesired. Most importantly, consumer (and generator) information
is centrally collected, resulting in data privacy issues and making
the grid vulnerable to intentional attacks, e.g. through hacking the
central computer. Recent cyber attacks \cite{NewYorkTimes2017,OliverWyman2017}
demonstrated that even large companies, like banks and logistic enterprises,
are vulnerable to such hacking attempts. Moreover, we do not fully
understand to date how and under which conditions central control
is feasible for such large, distributed and nonlinear system as power
grids; in particular, generally valid stability guarantees are missing
entirely. 

\revise{
 As an alternative to central control paradigms, decentralized control algorithms have been proposed, e.g. in  \cite{Bahrami2017,Mohammadi2018} or  \cite{Schweppe1982,Walter2014,Walter2016}.
Decentral control could work by providing local feedback to the consumer
or generator by evaluating local frequency deviations from the reference
frequency and providing a local control signal.} Thereby, it extends
the basic mechanism of \emph{primary control} \cite{Machowski2011}
to decentral generators and consumers by coupling frequency and price
incentives.

\revise{It remains
unclear whether and how  both central and decentral control actions are capable to coping with non-Gaussian fluctuations.
The viability of many control methods has only been tested using historical time series \cite{Margaris,Short} or simple stochastic models \cite{Mercier}.
The aforementioned decentral smart grid control has been demonstrated
to induce frequency-stable operation in small and moderately sized
grids with balanced demands and supplies that are static \cite{Schaefer2015,Schaefer2016}. Nevertheless,
how stochastic perturbations of the power are influenced by such a
control is not understood. }

To gain first fundamental insights about these questions, and see
whether non-Gaussian fluctuations may be limited at all by such decentral
control schemes, we here study the impact of decentral smart grid
control on power grid networks. We integrate external power fluctuations
as external noise sources and study how these transform to frequency
fluctuations when modeling the power grid as coupled (virtual) synchronous
machines.  We combine noise-driven grid modeling with stochastic
analysis of frequency data recorded in real transmission grids that
exhibit non-Gaussian frequency fluctuations. We consider both Gaussian
noise processes and non-Gaussian frequency measurements to extract
the power fluctuations, analyze and quantify how decentralized control
may reduce frequency fluctuations and may thus contribute to improve
dynamic stability of electricity grids.

\section{Modeling power grid frequency fluctuations}

We model power grid dynamics using a coarse-grained model to integrate
real frequency measurements with mathematical methods of stochastic
analysis. We aggregate several synchronous generators within one region
together with the regional loads into one (virtual) synchronous machine
\cite{Biegel2014}. An excess of generated power will make such a
machine an effective generator, while urban regions are modeled as
effective loads (consumers). Within a simple dynamical description
of coupled synchronous machines using the swing equation \cite{Machowski2011,Kundur1994},
each node $i\in\left\{ 1,...,N\right\} $ is modeled via
its voltage phase angle $\theta_{i}\left(t\right)$ and angular velocity
$\omega_{i}\left(t\right)$ as

\begin{eqnarray}
\frac{\text{d}}{\text{d}t}\theta_{i} & = & \omega_{i},\label{eq:Equation of motion power grid}\\
M_{i}\frac{\text{d}}{\text{d}t}\omega_{i} & = & F_{i}\left(\boldsymbol{\theta},\boldsymbol{\omega},t\right)+C_{i}\left(\boldsymbol{\theta},\boldsymbol{\omega},t\right),\nonumber 
\end{eqnarray}
with inertia $M_{i}$, intrinsic, potentially noisy dynamics, $F_{i}\left(\boldsymbol{\theta},\boldsymbol{\omega},t\right)$
and control $C_{i}\left(\boldsymbol{\theta},\boldsymbol{\omega},t\right)$.
Neglecting ohmic losses and assuming constant voltage amplitudes (well
justified for high voltage transmission grids \cite{Kundur1994,Filatrella2008}),
the intrinsic dynamic is given as 
\begin{equation}
F_{i}\left(\boldsymbol{\theta},\boldsymbol{\omega},t\right)=P_{i}^{{\rm in}}(t)-\kappa_{i}^{D}\omega_{i}+\sum_{j=1}^{N}K_{ij}\sin\left(\theta_{j}-\theta_{i}\right),
\end{equation}
where at each node we have active injected power $P_{i}^{{\rm in}}(t)$,
including fluctuations, damping $\kappa_{i}^{D}$ and coupling matrix
$K_{ij}$. For now, we analyze the impact of random perturbations
in the power $P_{i}^{{\rm in}}(t)$ in the absence of (decentral)
control, $C_{i}\equiv0$.
\revise{Throughout this manuscript, we use a per-unit notation so that all quantities have unit 1 or powers of $1/{\rm second}$, see e.g. \cite{Filatrella2008}}

\subsection{Gaussian noise}

As a simple model for power fluctuations, we consider white Gaussian
noise such that the active injected power 
\begin{equation}
P_{i}^{{\rm in}}(t)=P_{0,i}^{{\rm in}}+\sigma_{i}^{P}\Gamma_{i}\left(t\right)
\end{equation}
at node $i$ is a sum of a constant term $P_{0,i}^{{\rm in}}$ and
a noise term $\Gamma_{i}\left(t\right)$ multiplied by an noise amplitude
$\sigma_{i}^{P}$. Thereby, we allow different noise distributions at individual nodes and only require a time-average of zero for the noise.

\revise{We assume the damping $\kappa_{i}^{D}$ to be proportional to the inertia
at each node \cite{Weixelbraun2009} so that $\gamma=\kappa_{i}^{D}/M_{i}$
for all $i$. While this may not be true in all cases, this assumption is necessary to obtain analytical results.} Furthermore, we assume the power to be balanced on average,
i.e., $\sum_{i=1}^{N}P_{0,i}^{{\rm in}}=0$ and the coupling to be
symmetrical $K_{ij}=K_{ji}$. The dynamics of the bulk angular velocity
$\bar{\omega}=\sum_{i=1}^{N}\omega_{i}M_{i}/\sum_{i=1}^{N}M_{i}$
is then given by \revise{
\begin{equation}
\frac{\text{d}}{\text{d}t}\bar{\omega}=-\gamma\bar{\omega}+\frac{\sum_{i=1}^N\sigma_i^P\Gamma_i(t)}{\sum_{i=1}^NM_i}.\label{eq:bulk dynamics}
\end{equation} 
}

The sum of Gaussian processes is again a Gaussian process \cite{Gardiner1985,Samorodnitsky1994} 
such that \revise{
\begin{equation}
\frac{\sum_{i=1}^N\sigma_i^P\Gamma_i(t)}{\sum_{i=1}^NM_i}=\bar{\sigma}^{P}\bar{\Gamma},
\end{equation}}
where the aggregated noise amplitude $\bar{\sigma}^{P}$ is defined as\revise{
\begin{equation}
\bar{\sigma}^{P}=\frac{\sqrt{\sum_{i=1}^{N}\left(\sigma_{i}^{P}\right)^{2}}}{\sum_{i=1}^{N}M_{i}}.
\end{equation}}
To obtain the probability distribution
for the angular velocity $p\left(\bar{\omega}\right)$, we have to formulate and solve the Fokker-Planck equation of the dynamics (\ref{eq:bulk dynamics}) . The resulting distribution \cite{Schaefer2017a,Gardiner1985,Risken1984a} is (also) a Gaussian with  standard deviation 
\begin{equation}
\bar{\sigma}^{\omega}=\frac{\bar{\sigma}^{P}}{\sqrt{2\gamma}}.\label{eq:Predicted standard deviation}
\end{equation}
A final quantity of interest is the autocorrelation $c\left(\Delta t\right)$
as a function of the time lag $\Delta t$, which for process (\ref{eq:bulk dynamics})
is a decaying exponential \cite{Schaefer2017a,Gardiner1985}
\begin{equation}
c\left(\Delta t\right)=\exp\left(-\gamma\Delta t\right).\label{eq:Autocorrelation decay}
\end{equation}
With these results, we now analyze frequency recordings from a real
power grid, assuming that we measure the bulk frequency $\bar{f}$
and convert it to the bulk angular velocity via $\bar{\omega}=2\pi\left(\bar{f}-f_{R}\right)$.
Given the statistics for $\bar{\omega}$, we estimate its standard
deviation $\bar{\sigma}^{\omega}$ and compute the autocorrelation $c\left(\Delta t\right)$
to determine the damping to inertia ratio $\gamma$ from an exponential
fit. 

\begin{figure*}
\begin{centering}
\includegraphics[width=1.9\columnwidth]{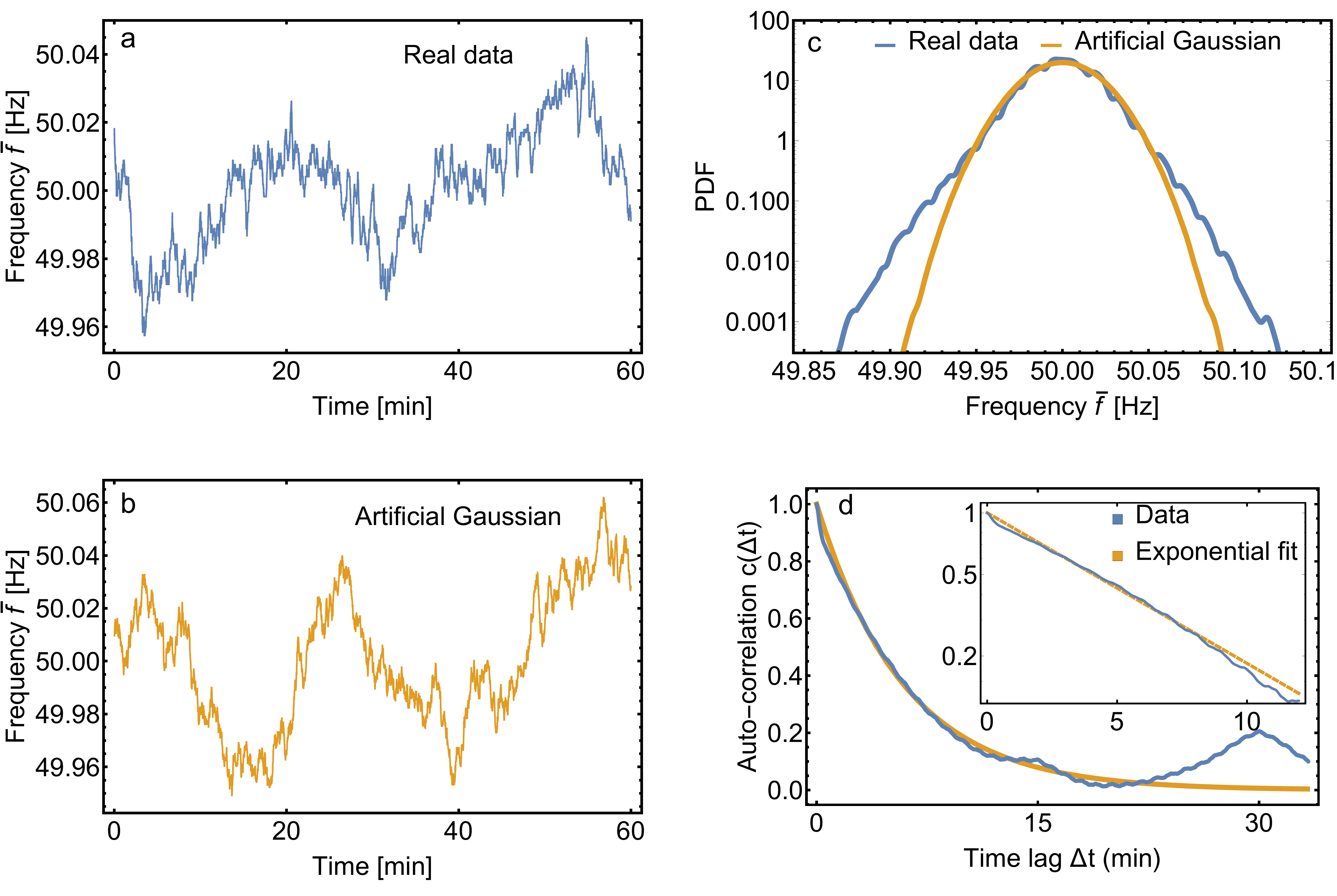}
\par\end{centering}
\caption{Real power grid frequency fluctuations are not Gaussian and show an exponential decay of the autocorrelation. a: Sample
trajectory of the real frequency trajectory. b: Sample trajectory
of an artificial Gaussian frequency trajectory, assuming an underlying
Ornstein-Uhlenbeck process \eqref{eq:bulk dynamics}. c: The probability density function (PDF)
of the bulk frequency $\bar{f}$ comparing real data with artificial
Gaussian data. d: The autocorrelation function of the real frequency
data decays approximately according an exponential fit. The inset uses a log-linear plot to highlight the exponential decay. In addition, we observe regular correlation peaks, associated with trading \cite{Schaefer2017a}. The real data
has much heavier tails than a Gaussian distribution would predict
(Kurtosis of approx. $3.8$ instead of $3$ for Gaussian distribution). We use data by \emph{50Hertz} from 2015 with
a 1 second resolution describing the Continental European power grid
\cite{50Hertz-UCTE2016}. \label{fig:Frequency-trajectory-and Histogram Real vs aritfical}}
\end{figure*}

We compare real frequency measurements with our theory and a naive
Gaussian assumption in Figure \ref{fig:Frequency-trajectory-and Histogram Real vs aritfical}.
Panels a,b illustrate an artificial trajectory following such
an Ornstein-Uhlenbeck process, as given by Eq. (\ref{eq:bulk dynamics})
and a real frequency time series recorded in 2015 in the Continental
European Grid \cite{50Hertz-UCTE2016}. Comparing the statistics systematically,
we find 
that the real frequency data exhibit a higher likelihood of large
fluctuations than that predicted by the best fitting Gaussian distribution (panel c).
To model these heavy tails, we thus generalize Gaussian to arbitrary
stable distributions \cite{Samorodnitsky1994} to characterize the
frequency deviations. Finally, panel d of Fig. \ref{fig:Frequency-trajectory-and Histogram Real vs aritfical}
shows the exponential decay of the autocorrelation function predicted
by Eq. (\ref{eq:Autocorrelation decay}). 

\subsection{Extracting power fluctuations from frequency fluctuations}

Given real frequency measurements, we extract the statistics of the
aggregated power fluctuations following a generalized Fokker-Planck
equation \cite{Denisov2009}, i.e. we still use Eq. (\ref{eq:bulk dynamics})
but allow the noise $\Gamma_{i}\left(t\right)$ to follow non-Gaussian
distributions \cite{Metzler2000}. Let us denote the characteristic
equation, i.e., the Fourier transform of the probability density function,
of the collective power noise as $S^{P}\left(k\right)$. Then, the
following relations with respect to the characteristic function of
the angular velocity $S^{\omega}\left(k\right)$ hold \cite{Schaefer2017a}:

\begin{equation}
S^{\omega}\left(k\right)=\exp\left[\frac{1}{\gamma}\int_{0}^{k}\frac{\ln\left(S^{P}\left(z\right)\right)}{z}\text{d}z\right],
\end{equation}

\begin{equation}
S^{P}\left(k\right)=\exp\left[\gamma k\frac{\partial}{\partial k}\ln\left(S^{\omega}\left(k\right)\right)\right],\label{eq:Input distribution from final distribution}
\end{equation}
similar to the Gaussian case, individual nodes might have different
distributions, but we require the average of the perturbation to be
zero and it has to be possible to aggregate all contributions. 

\begin{figure*}
\begin{centering}
\includegraphics[width=1.9\columnwidth]{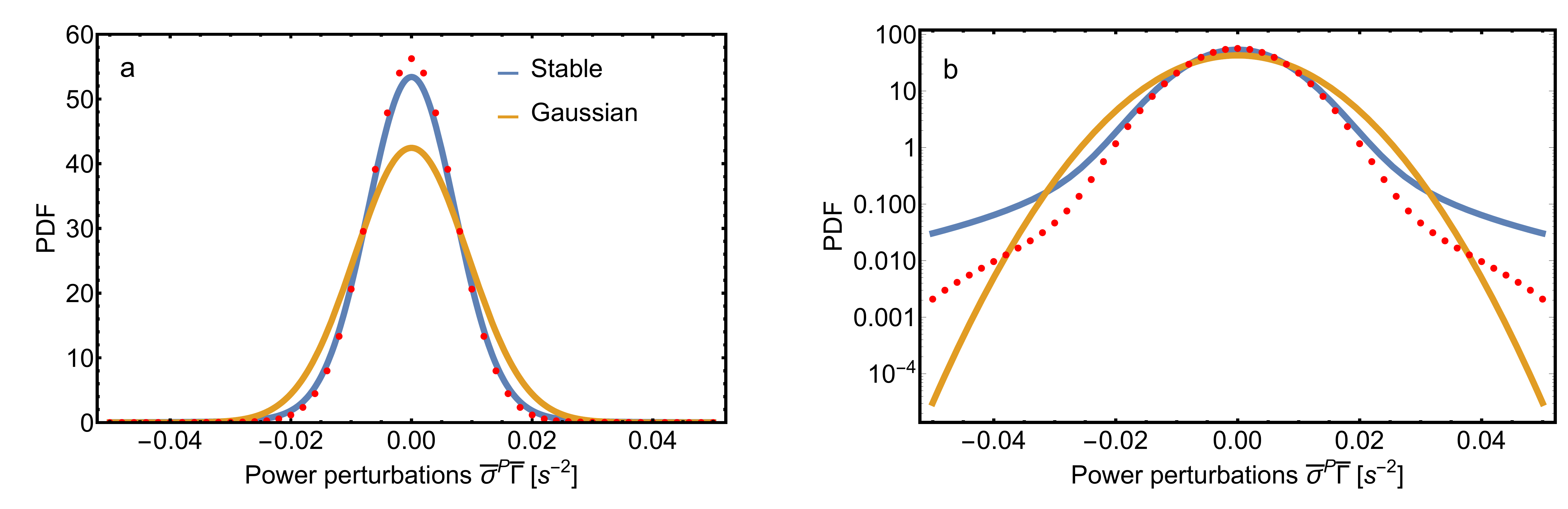}
\par\end{centering}
\caption{Using frequency measurements, we extract approximate power fluctuations.
 We plot the probability density function (PDF) of the estimated power fluctuations,
based on frequency measurements, using Eq. (\ref{eq:Input distribution from final distribution})
and assuming noise following the best fitting Gaussian or non-Gaussian stable distribution
(solid curves). The stable noise distribution is a more likely description
for our data (red dots), only overestimating the tails slightly. a: Linear scale for the PDF. b: Log-scale for the PDF.  We use data by \emph{50Hertz} from 2015 with a 1 second
resolution describing the Continental European power grid \cite{50Hertz-UCTE2016}.
\label{fig:PowerStatisticsExtracted from frequency statistics}}
\end{figure*}

Using measurements from the continental European grid \cite{50Hertz-UCTE2016},
we first perform a maximum likelihood analysis \cite{Bohm2010}
to identify Gaussian and the more general stable distributions as
generally good fits to the data. Next, we extract the approximate characteristic
function of the angular velocity $S^{\omega}\left(k\right)$ from
the distribution and apply Eq. (\ref{eq:Input distribution from final distribution})
to extract the distribution of the aggregated power fluctuations $S^{P}\left(k\right)$.
Fig. \ref{fig:PowerStatisticsExtracted from frequency statistics}
displays the results when using the data directly (red dots), compared
to the assumption of Gaussian noise (orange curve) and non-Gaussian
stable noise (blue curve). Gaussian distributions tend to underestimate
the tails of the distribution (see Fig. \ref{fig:Frequency-trajectory-and Histogram Real vs aritfical})
so that we use stable distributions to complement our analysis. Stable
distributions (also known as $\alpha$-stable or Lévy-stable) are
characterized by a stability parameter $\alpha_{S}$, determining
the heavy tails, a skewness parameter $\beta_{S}$, a scale parameter
$\sigma_{S}$, which fulfills a similar role as the standard deviation
in the Gaussian case, and a location parameter $\mu_{S}$. \revise{In the
remainder of the paper we assume that the noise follows a stable distribution,
simplifying it by assuming centered and symmetric distributions, i.e.
$\beta_{S}=\mu_{S}=0$ for all nodes. Furthermore, we assume that the stability parameter $\alpha_{S}$ is identical at all nodes, thereby ensuring that the resulting bulk distribution is a stable distribution, as observed in the data.} Stable distributions contain Gaussian distributions
as a special case given by the stability parameter $\alpha_{S}=2$.
In the plots below, we use the term stable distributions only in connection
with $\alpha_{S}<2$, i.e. non-Gaussian stable distributions. 

To aggregate several independent stable distributions, we require
that the noise at each node has identical stability parameter $\alpha_{S}$
but it may have an arbitrary scale parameter $\sigma_{S}$. Let $\sqrt{2} \sigma_{S,i}^{P}$
be the scale parameter of the stable power perturbations at node $i$, where the factor $\sqrt{2}$ is necessary to reproduce the Gaussian case for $\alpha_S=2$.
Then, the scale parameter of the bulk angular velocity is given as
\cite{Schaefer2017a,Denisov2009} 
\begin{equation}
\bar{\sigma}_{S}^{\omega}=\frac{1}{\sqrt{2}\sum_{i=1}^{N}M_{i}}\left[\frac{1}{\gamma\alpha_{S}}\sum_{i=1}^{N}\left(\sigma_{S,i}^{P}\right)^{\alpha_{S}}\right]^{1/\alpha_{S}}.\label{eq:Predicted scale parameter}
\end{equation}

\section{Decentral control}

\begin{figure}
\begin{centering}
\includegraphics[width=0.99\linewidth]{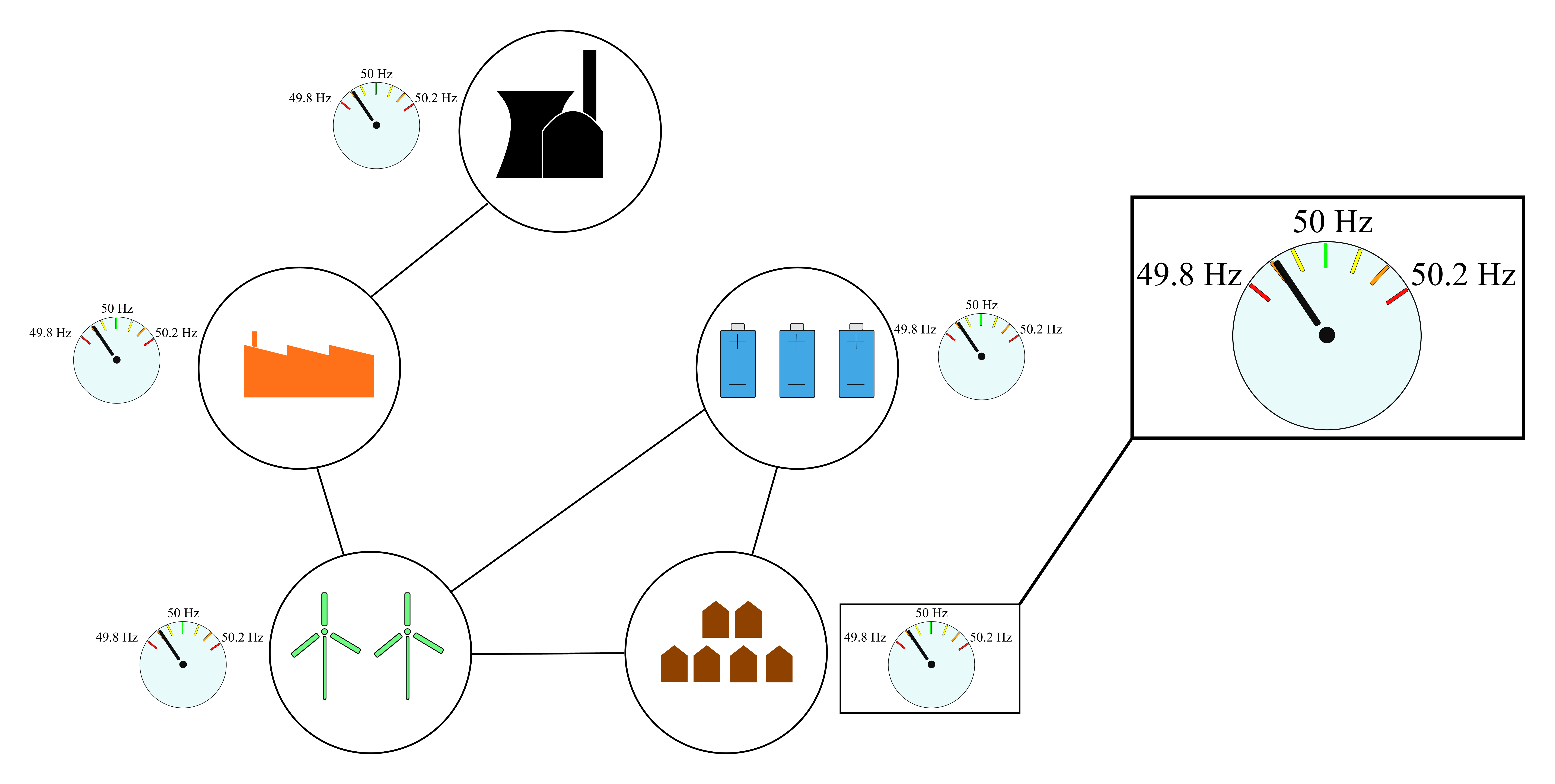}
\par\end{centering}
\caption{Decentral Smart Grid Control uses local frequency measurements at
all nodes, regardless of whether they are consumers, generators or storage
facilities to stabilize the grid. Driven by price incentives, customers
should increase generation (reduce consumption) in case of low frequencies
and decrease generation (increase consumption) in case of high frequencies.
\label{fig:Decentral-Smart-Grid-Control}}
\end{figure}

In the previous section, we have explored how to extract the statistics
of power fluctuations from power grid frequency measurement data,
inferring the underlying distribution of power disturbances. These
considerations assumed that there is no control $C_{i}$ at a given
node. We follow the proposal of Decentral Smart Grid Control, introduced
in \cite{Walter2014,Walter2016} and mathematically modeled in \cite{Schaefer2015,Schaefer2016}
to motivate an additional control term for the swing equation (\ref{eq:Equation of motion power grid})
as follows. All grid participants use the local grid frequency to
determine an energy shortage (in case of low frequencies) or energy
abundance (in case of high frequencies). Based on price incentives,
customers then adapt their consumption and generation to keep the
grid closer to the desired frequency, see Fig. \ref{fig:Decentral-Smart-Grid-Control}
for an illustration.

Let us assume that the injected power $P_{i}^{{\rm in}}(t)$ at node
$i$ is given as the difference of supply $\mathcal{S}_{i}(p_{i})$
and demand $\mathcal{D}_{i}(p_{i})$, which both depend on the price
$p_{i}$
\begin{equation}
P_{i}^{{\rm in}}(t)=\mathcal{S}_{i}(p_{i})-\mathcal{D}_{i}(p_{i}).
\end{equation}
We expect supply to increase and demand to decrease with increasing
prices so that overall, $P_{i}^{{\rm in}}(p_{i})$ increases with
increasing price $p_{i}$. For simplicity, we assume this dependency
to be linear for prices close to the equilibrium
\begin{equation}
P_{i}^{{\rm in}}(p_{i})=P_{i,0}^{{\rm in}}+c_{i}^{(1)}\left(p_{i}-p_{i,0}\right),\label{eq:Power as function of price}
\end{equation}
with equilibrium injected power $P_{0,i}^{{\rm in}}$, equilibrium
price $p_{i,0}$ and price-dependency $c_{i}^{(1)}$. Next, we need
to determine how the price adapts with respect to the frequency. Following
recent proposals \cite{Walter2014,Walter2016}, we will assume that
the price is a linear decreasing function of the frequency $f_{i}$,
given in terms of the angular velocity $\omega_{i}=2\pi (f_{i}-f_R)$ as
\begin{equation}
p_{i}(\omega_{i})=p_{i,0}-c_{i}^{(2)}\omega_{i},\label{eq:Price as a function of the frequency}
\end{equation}
with price constant $c_{i}^{(2)}$. Thereby, the injected power becomes
a linear function of the angular velocity as 
\begin{equation}
P_{i}^{{\rm in}}(\omega_{i})=P_{0,i}^{{\rm in}}-\kappa_{i}^{C}\omega_{i},\label{eq:frequency-dependent power}
\end{equation}
with $\kappa_{i}^{C}=c_{i}^{(1)}c_{i}^{(2)}$, i.e., it is a product
of the price-dependency of the node $i$ and the slope of the price-frequency
curve\@. Using the notation of Eq. (\ref{eq:Equation of motion power grid}),
we formulate our control term as
\begin{equation}
C_{i}\left(\boldsymbol{\theta},\boldsymbol{\omega},t\right)=-\kappa_{i}^{C}\omega_{i}.\label{eq:Final Control term}
\end{equation}
Inspecting (\ref{eq:Final Control term}), we notice that adding our
decentralized control adds effective damping to the grid, similar
to control applied on generator sides today \cite{Sun2010}. The main
benefit is that we generalize the notion of primary control and make
it applicable to all grid participants, generators, consumers as well
as other services that e.g. provide electricity storage.

\section{Reducing fluctuation risks with decentral control}

\begin{figure}
\begin{centering}
\includegraphics[width=0.95\columnwidth]{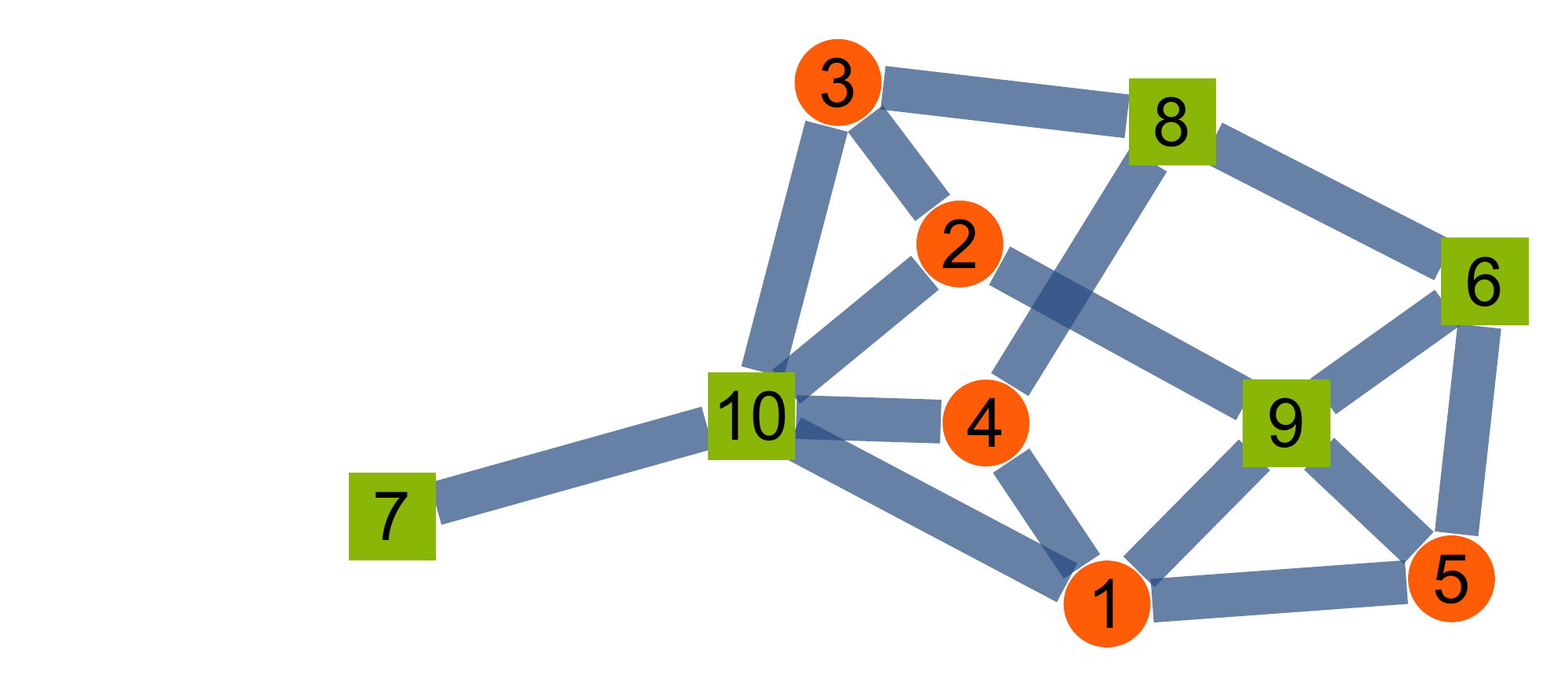}
\par\end{centering}
\caption{\revise{Ten node test grid used for the simulations.
Red circles (indices 1--5) represent consumers with $P^{{\rm in}}({\rm  Con})=-1s^{-2}$ and green squares (indices 6--10) represent generators with $P^{{\rm in}}({\rm Gen})=+1s^{-2}$.
If not stated otherwise, the inertia values are $M_{i}=(1.1,1.7,7,8.7,3.2,9.8,0.7,5.8,0.2,0.9)$
which was obtained by randomly drawing an inertia value in the interval
$M_{i}\in\left[0.1,10\right]$ for each node. The coupling matrix is
$K_{ij}=0$ if two nodes are not connected and 
$K_{ij}=4s^{-2}$ otherwise.} \label{fig:Artificial-ten-node test grid}}
\end{figure}

Does a decentralized control $C_{i}$, as proposed above, reduce fluctuation
risks? We investigate this question using a ten node test grid, see
Fig. \ref{fig:Artificial-ten-node test grid}, considering different
control settings, using Gaussian and non-Gaussian stable noise with stability
parameter $\alpha_{S}=1.5$. \revise{For each setting, we evolve the stochastic
differential equation given by Eq. (\ref{eq:Equation of motion power grid})
and compare it to the analytical prediction of the  bulk frequency distribution $p\left(\bar{\omega}\right)$, based on the theory from the last section and \cite{Schaefer2017a}. We vary the strength of control applied via a control parameter $\kappa^C$.}
For our analysis restricted to Gaussian noise, we extract the standard
deviation of the distribution and compare it to the predicted standard
deviation in Eq. (\ref{eq:Predicted standard deviation}). Similarly,
for non-Gaussian stable noise, we extract the scale parameter and
compare it to the predicted one in Eq. (\ref{eq:Predicted scale parameter}). \revise{All simulations use $\sigma_S=1/100$ at all nodes, i.e. for the Gaussian case, we use $\sigma=\sqrt{2}/100$. Heterogeneous noise does produce similar results (not shown).}

\paragraph{Homogeneous control to inertia}

\begin{figure*}
\begin{centering}
\includegraphics[width=1.9\columnwidth]{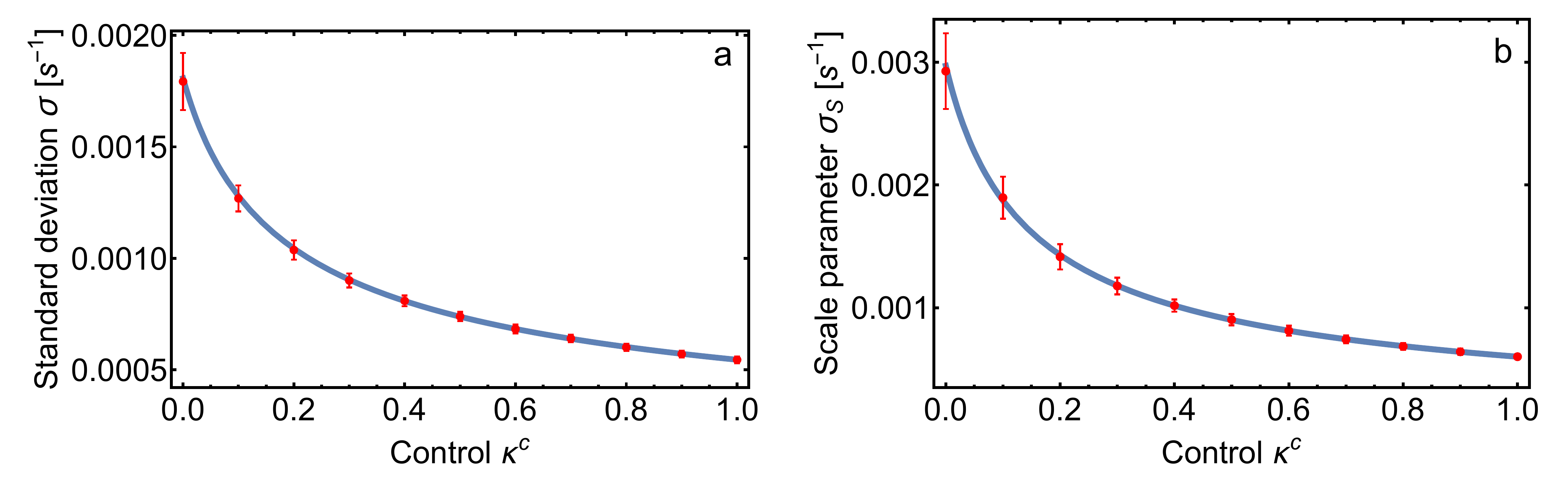}
\par\end{centering}
\caption{Fluctuations are reduced by control as predicted by the theory. a:
We plot the predicted standard deviation $\sigma$, based on (\ref{eq:Predicted standard deviation})
versus simulation results. b: We plot the predicted scale parameter
$\sigma_{S}$, based on (\ref{eq:Predicted scale parameter}), versus simulation results. Simulations used
the ten node network, see Fig. \ref{fig:Artificial-ten-node test grid},
with both damping $\kappa_{i}^{D}$ and control $\kappa_{i}^{C}$
proportional to the inertia $M_{i}$ with $\kappa_{i}^{D}/M_{i}=0.1s^{-1}$
and $\kappa_{i}^{C}=10\kappa^{C}\kappa_{i}^{D}$. The dots give the
mean value based on 100 independent experiments of independent random
realizations of the noise trajectory. For each trajectory, we evaluate
1000 data points. The error bars show the standard deviation between
these 100 runs. \label{fig:Homogeneous ratio}}
\end{figure*}

\revise{If we assume the grid to have a homogeneous damping to inertia ratio
$\kappa_{i}^{D}/M_{i}=\gamma=0.1s^{-1}$, we may use equations (\ref{eq:Predicted standard deviation})
and (\ref{eq:Predicted scale parameter}) for Gaussian and non-Gaussian
noise respectively. Furthermore, if the control $\kappa_{i}^{C}$
at each node is also proportional to the inertia $\kappa_{i}^{C}=10\kappa^{C}\kappa_{i}^{D}$,
we have a precise prediction of how the standard deviation
and scale parameter depend on the control. Fig. \ref{fig:Homogeneous ratio} shows the great
agreement of theory and simulations. Namely, increasing control $\kappa^{C}$
decreases the scale parameter (or standard deviation) and thereby
reduces fluctuation risks.}

\paragraph{Heterogeneous control to inertia}

\begin{figure*}
\begin{centering}
\includegraphics[width=1.9\columnwidth]{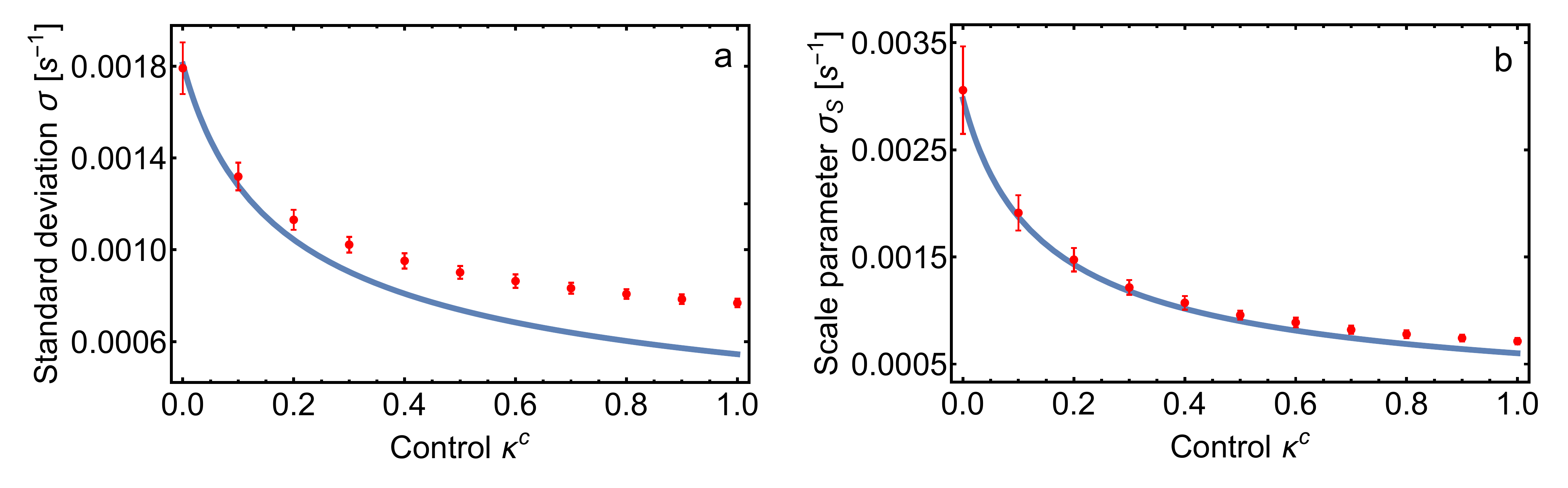}
\par\end{centering}
\caption{Control also reduces fluctuations when using heterogeneous damping
to inertia ratios. a: We plot the predicted standard deviation $\sigma$,
based on (\ref{eq:Predicted standard deviation}) versus simulation
results. b: We plot the predicted scale parameter $\sigma_{S}$, based on (\ref{eq:Predicted scale parameter}), 
versus simulation results. Simulations used the ten node network,
see Fig. \ref{fig:Artificial-ten-node test grid}, with the average
damping being proportional to the average inertia $\sum_{i=1}^{N}\kappa_{i}^{D}/\sum_{i=1}^{N}M_{i}=0.1s^{-1}$
and control $\kappa_{i}^{C}$ proportional to the individual damping
$\kappa_{i}^{C}=10\kappa^{C}\kappa_{i}^{D}$. The prediction overestimates
the effect of control. We use the mean damping and mean inertia for
the analytical prediction (blue curves). The damping realization for
this plot is $\kappa_{i}^{D}=(0.306,0.494,0.158,0.188,0.573,0.089,0.592,0.849,0.425,0.236)s^{-1}$.
\label{fig:Heterogeneous ratio} The dots give the mean value based
on 100 independent experiments of independent random realizations
of the noise trajectory. For each trajectory, we evaluate 1000 data
points. The error bars show the standard deviation between these 100
runs.}
\end{figure*}

\revise{Next, let us drop the assumption that the damping $\kappa_{i}^{D}$ is proportional to the inertia
$M_{i}$. Instead, we determine damping values $\kappa_{i}^{D}$  so that averaged over all nodes the total ratio is the same
as before $\sum_{i=1}^{N}\kappa_{i}^{D}/\sum_{i=1}^{N}M_{i}=0.1s^{-1}$.
The control at each node is still proportional to the damping with $\kappa_{i}^{C}=10\kappa^{C}\kappa_{i}^{D}$.
 Inspecting Fig. \ref{fig:Heterogeneous ratio} reveals that increasing
the control $\kappa^{C}$ still decreases the observed width of the
distribution, i.e. standard deviation $\sigma$ and scale parameter $\sigma_S$ decrease. However, the simulations do no longer perfectly align
with the theoretical predictions. Instead, the simulations tend to display a wider distribution than expected based on a homogeneous damping to inertia ratio.  These deviations
between simulations and theory appear to be smaller for non-Gaussian
noise. Overall, while increasing control reduces fluctuations, heterogeneous damping and inertia values lead to larger fluctuations than homogeneous values.}

\begin{figure*}
\begin{centering}
\includegraphics[width=1.9\columnwidth]{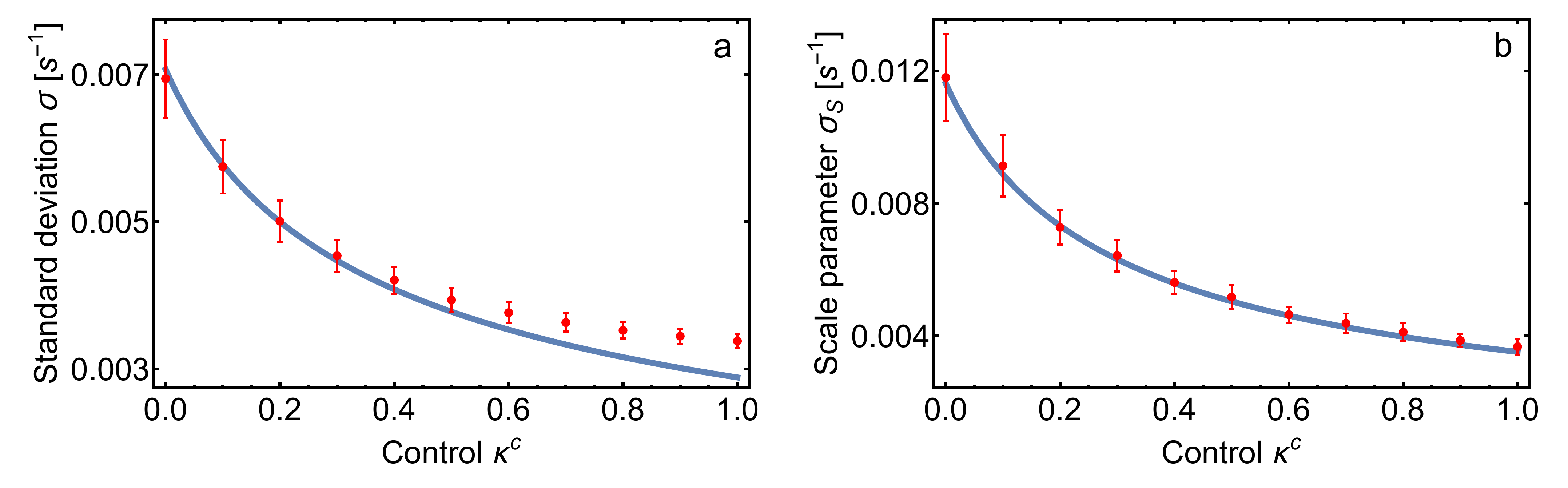}
\par\end{centering}
\caption{Control also reduces fluctuations when only controlling generators.
a: We plot the predicted standard deviation $\sigma$, based on (\ref{eq:Predicted standard deviation})
versus simulation results. b: We plot the predicted scale parameter
$\sigma_{S}$, based on (\ref{eq:Predicted scale parameter}), versus simulation results. Simulations used
the ten node network, see Fig. \ref{fig:Artificial-ten-node test grid},
with identical inertia $M_{i}=M=1$ and damping $\kappa_{i}^{D}=\kappa^{D}=0.1s^{-1}$
for all nodes. The control is proportional to the inertia but only
for generators, i.e., for nodes with $P^{{\rm in}}_{i}>0$, the control is $\kappa_{i}^{C}=10\kappa^{C}\kappa_{i}^{D}$ and zero otherwise.
The prediction overestimates the effect of control slightly. We use
half the value of $\kappa^{C}$ for our analytical prediction (blue
curves).\label{fig:IdenticalMachines ControlGen} The dots give the
mean value based on 100 independent experiments of independent random
realizations of the noise trajectory. For each trajectory, we evaluate
1000 data points. The error bars show the standard deviation between
these 100 runs.}
\end{figure*}

\paragraph{Controlling generators only}
\revise{
Often, control is assumed to be a task to be fulfilled primarily by
the generators and not by the consumers \cite{Machowski2011,Wood2013}.
Here, we implement this control paradigm using identical machines
with $M_{i}\equiv M=1$ for all $i$ and homogeneous damping $\kappa_{i}^{D}=\kappa^{D}=0.1s^{-1}$.
The control is $\kappa_{i}^{C}=10\kappa^{C}\kappa_{i}^{D}$ for generators,
i.e. for nodes with $P^{{\rm in}}_{i}>0$, and zero otherwise. To match the simulation,
we include only $\kappa^{C}/2$ in our predictions, since only half
of the network is controlled, and observe a good match between
theory and simulation in Fig. \ref{fig:IdenticalMachines ControlGen}.
Applying frequency-dependent control only at the generator nodes already decreases  the overall frequency fluctuations in the network. Interestingly, theory and simulation agree very well in this case, although the theory was derived assuming homogeneous parameters.}

\section{Discussion}

\revise{Overall, we have shown that non-Gaussian effects are present in real
power grid frequency statistics, with non-Gaussian stable distributions
as good fits (Fig. \ref{fig:Frequency-trajectory-and Histogram Real vs aritfical}).
Furthermore, we have made some earlier results \cite{Schaefer2017a}
more explicit, for example when extracting the distribution of power
fluctuations from pure frequency measurements (Fig. \ref{fig:PowerStatisticsExtracted from frequency statistics}).
To reduce the effect of these power fluctuations on the grid frequency, a decentralized control
paradigm could be used that relies on local grid frequency measurements
and price incentives to effectively provide additional droop control
at all nodes (Fig. \ref{fig:Decentral-Smart-Grid-Control}). Such
a control intrinsically avoids privacy issues or vulnerability concerns
as no communication infrastructure is necessary. Using both simulations
and stochastic analysis, we have predicted the effectiveness of such a decentralized
control, extending previous results on deterministic stability \cite{Schaefer2015,Schaefer2016} to stochastic stability.
In particular, we have derived a scaling of the general scale parameter in equation (\ref{eq:Predicted scale parameter}).
Investigating a ten node test grid, we found that our stochastic
analysis matches the simulations results for the standard deviation
(or scale parameter in the case of non-Gaussian stable noise) very
well (Fig. \ref{fig:Homogeneous ratio}). Increasing decentral control
reduces fluctuation risks by decreasing standard deviation (or scale
parameter) of the resulting frequency distribution. Even for heterogeneous damping to inertia ratios, we observe a reduction
of fluctuations. However, fluctuation risks are reduced less efficiently
in the presence of heterogeneous ratios when compared to the predictions
and the precise scaling of the fluctuations is only approximately
described by our theory (Fig. \ref{fig:Heterogeneous ratio}). Similarly,
controlling only parts of the network, e.g. because consumers do not
participate in demand control schemes, still reduces fluctuation risks,
approximately as predicted by the theory (Fig \ref{fig:IdenticalMachines ControlGen}).
}

% for control that
%is proportional to the inverse inertia \cite{Weixelbraun2009}.

\revise{Our analysis used several necessary simplifying assumptions. Most crucially, we assumed the inertia to be proportional to the damping at each node. This assumption is crucial since thereby the stochastic equation of motion Eq. \eqref{eq:bulk dynamics} becomes a linear one-dimensional equation. Otherwise, we would end with a high-dimensional, nonlinear Fokker-Planck equation for the bulk frequency $p(\bar\omega)$. Since Fokker-Planck equations are partial differential equations, solving such a complex equation analytically becomes impossible \cite{Gardiner1985}. Instead, we made the above-mentioned simplifications to also include the  non-Gaussian effects observable in real power grid dynamics.}

In conclusion, decentral smart grid control may be capable of
reducing frequency fluctuation in power grids which exhibit Gaussian
or more general non-Gaussian fluctuations of the grid frequency. We
formulated an approximate stochastic theory to predict the effectiveness
of control for a family of noise distributions and supported these
predictions with simulations for Gaussian and stable noise. Based
on all performed simulations (not all shown), the precise network
topology and distribution of power seem negligible, as long as lines
are not heavily loaded (the coupling $K_{ij}$ is large compared to
the power $P^{{\rm in}}_{i}$). The main benefit of the decentralized control
scheme is its inclusion of all grid participants, thereby distributing
the control burden throughout the network.

Future research may focus on open theoretical and practical questions.
First, it remains unclear how individual node frequency statistics
are affected by Gaussian and non-Gaussian noise, respectively. Second,
there is no explanation yet for why our approximation seems to more
closely resemble the simulation data for non-Gaussian noise than it
does for Gaussian noise (see Fig. \ref{fig:Homogeneous ratio} b).
Third, control actions may not always be instantaneous but instead
could follow a delayed signal, compare for instance \cite{Schaefer2015,Schaefer2016,Milano2012}.
How such a delay affects the effectiveness of decentral control of fluctuations remains an open question as well.

\section*{ACKNOWLEDGMENT}

This work is supported through the German Science Foundation (DFG)
by a grant toward the Cluster of Excellence \textquotedblleft Center
for Advancing Electronics Dresden\textquotedblright{} (cfaed), the
Helmholtz Association (via the joint initiative \textquotedblleft Energy
System 2050\textemdash a Contribution of the Research Field Energy\textquotedblright{}
and grant no. VH-NG-1025) and the Federal Ministry of Education and
Research (BMBF Grant No 03SF0472F).

%%%%%%%%%%%%%%%%%%%%%%%%%%%%%%%%%%%%%%%%%%%%%%%%%%%%%%%%%%%%%%%%%%%%%%%%%%%%%%%%

%%%%%%%%%%%%%%%%%%%%%%%%%%%%%%%%%%%%%%%%%%%%%%%%%%%%%%%%%%%%%%%%%%%%%%%%%%%%%%%%
\providecommand{\url}[1]{#1}
\csname url@samestyle\endcsname

\begin{biography}[{\includegraphics[width=1in,height=1.25in,clip,keepaspectratio]{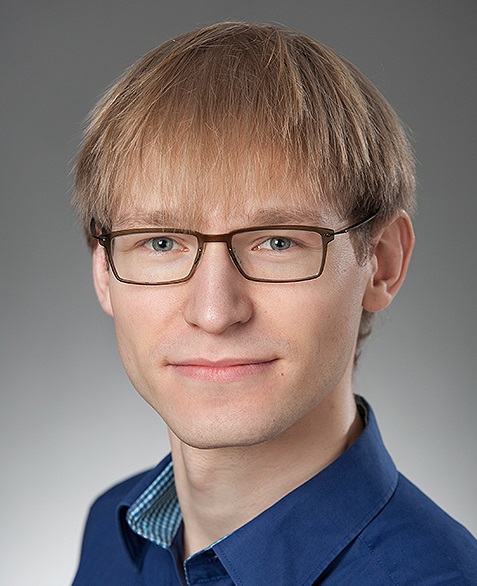}}] {Benjamin Sch\"afer}  received his Diplom degree in Physics from the University of Magdeburg, Germany in 2013. Pursuing his Ph.D. in G\"ottingen (Germany) London (United Kingdom) and Tokyo (Japan), he received his Ph.D. degree in physics in 2017 from the University of G\"ottingen. Since 2017 he is Postdoctoral Researcher at the Max Planck Institute for Dynamics and Self-Organization, G\"ottingen, Germany and the Technical University Dresden, Germany.  \end{biography}\begin{biography}[{\includegraphics[width=1in,height=1.25in,clip,keepaspectratio]{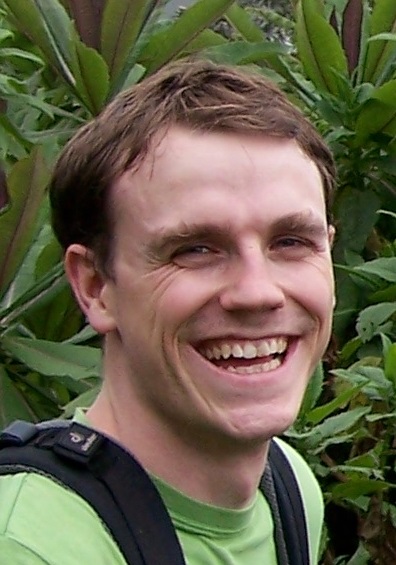}}] {Dirk Witthaut}  
received his Diploma (MSc) and PhD in Physics from the Technical University of Kaiserslautern, Germany, in 2004 and 2007, respectively. He has been a Postdoctoral Researcher at the Niels Bohr Institute in Copenhagen, Denmark and at the Max Planck Institute for Dynamics and Self-Organization in G\"ottingen, Germany and a Guest Lecturer at the Kigali Institute for Science and Technology in Rwanda. Currently, he is leading a Research Group at Forschungszentrum J\"ulich, Germany and he is a junior professor at the University of Cologne.  \end{biography}\begin{biography}[{\includegraphics[width=1in,height=1.25in,clip,keepaspectratio]{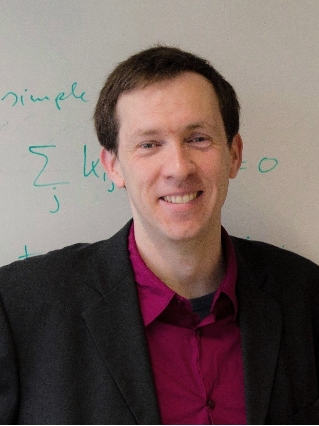}}] {Marc Timme} 
studied Physics and Mathematics at the University of W\"urzburg, Germany, and the State University of New York at Stony Brook, USA. He gained his doctorate in Theoretical Physics at the University of G\"ottingen. After two research stays at the Max Planck Institute for Flow Research and the Center for Applied Mathematics, Cornell University (USA), he was appointed to establish and head a broadly cross-disciplinary research group on Network Dynamics at the Max Planck Institute for Dynamics and Self-Organization. Marc Timme became an Adjunct Professor at the Institute for Nonlinear Dynamics at the University of G\"ottingen and is Co-Chair of the Association of Socio-Economic Physics of German Physical Society. He was Visiting Professor at TU Darmstadt, Germany, visiting faculty at the ETH Zurich Risk Center (Switzerland) and recently assumed a Chair for Network Dynamics as a Strategic Professor of the Cluster of Excellence Center for Advancing Electronics Dresden (cfaed) at TU Dresden, Germany. He received a research award of the Berliner Ungewitter Foundation, the Otto Hahn Medal of the Max Planck Society, and a Research Fellowship of the National Research Center of Italy.  \end{biography}

\end{document}